\documentclass[prl,twocolumn,showpacs,%superscriptaddress,
groupeaddress,preprintnumbers,amsmath,amssymb,tightenlines]{revtex4}
\usepackage{graphicx}
\newcommand{\beq}{\begin{equation}}
\newcommand{\eeq}{\end{equation}}
\newcommand{\bea}{\begin{eqnarray}}
\newcommand{\eea}{\end{eqnarray}}
\newcommand{\ba}{\begin{array}}
\newcommand{\ea}{\end{array}}
\newcommand{\bc}{\begin{center}}
\newcommand{\ec}{\end{center}}
\newcommand{\bml}{\begin{subequations}}
\newcommand{\eml}{\end{subequations}}
\newcommand{\commentout}[1]{{}}
\newcommand{\bk}{{\bf k}}

\newcommand{\adag}{a^\dagger}

\newcommand{\bdag}{b^\dagger}

\newcommand{\half}{\hbox{$\frac{1}{2}$}}

\newcommand{\Hc}{{\rm H.c.}}
\newcommand{\eq}[1]{(\ref{#1})}
\newcommand{\etal} {{\it et al.\/}}

\newcommand{\ibid} {{\it ibid. \/}}
\newcommand{\vol}[1]{{\bf #1}}
\newcommand{\comment}[1]{{}}

\newcommand{\cs}{$^{133}$Cs }
\newcommand{\csm}{$^{133}{\rm Cs}_2$ }

\begin{document}
\title{Coherent Population Trapping in a Feshbach-Resonant $^{133}$Cs Condensate}
\author{Matt Mackie}
\altaffiliation{This work was begun at QUANTOP--Danish National Research Foundation
Center for Quantum Optics, Department of Physics and Astronomy,
University of Aarhus, DK-8000 Aarhus C, Denmark.}
\affiliation{Department of Physics, Temple University, Philadelphia, PA 19122}
\date{\today}

\begin{abstract}
Recent experiments with Feshbach-resonant $^{133}$Cs Bose-Einstein condensates have led
to unexplained molecule formation: a sudden switch of the magnetic field to its resonance
value, followed by a finite hold time and another sudden switch to magnetic field values below
threshold, converts about a third of the initial condensate atoms into molecules.  Based on a
model of coherent conversion between an atomic condensate,  a molecular condensate, and
magnetodissociated noncondensate atom pairs of equal and opposite momentum, we find that
population trapping is strongly implicated as the physical mechanism responsible for molecule
formation in switch experiments.
\end{abstract}

\maketitle

{\em Introduction.}--Magnetoassociation creates a molecule from a pair of
colliding atoms when one of the atoms spin flips in the presence of a magnetic field
tuned near a Feshbach resonance~\cite{STW76}. Initial experiments with magnetoassociation
of Bose-Einstein condensates (BECs) led to dramatic losses of condensate atoms in the
neighborhood of resonance~\cite{INO98}, a collapsing condensate with a burst of atoms
emanating from the remnant condensate~\cite{DON01}, increased losses for decreasing
interaction times~\cite{CLA02}, and coherent oscillations between remnant and burst
atoms~\cite{DON02}. Whereas atom-molecule coherence is both necessary and sufficient to
the explain these observations, the phase space densities are insufficient to warrant a
claim to molecular condensate (MBEC) formation; hence, subsequent experiments focused on
magnetic-field sweeps across the Feshbach resonance~\cite{DUE04}, which led to the
unambiguous observation of BEC-MBEC conversion~\cite{KXU03}.

The backbone of these experiments is coherent evolution
between an atomic condensate, a molecular condensate condensate, and magnetodissociated
noncondensate atoms pairs of equal and opposite momentum~\cite{MAC02,KOK02}. In
particular, so-called rapid adiabatic passage arises because the ground state of the Feshbach
system is all atoms far above the molecular-dissociation threshold and all molecules far below
it, so that a slow sweep of the magnetic field from one extreme to the other converts atoms into
diatomic molecules. The colloquial understanding is that a magnetic field sweep directed from
below to above threshold converts an initial atomic condensate into noncondensate atom pairs of
equal and opposite momentum, since molecular dissociation is energetically favored above
threshold, whereas a sweep directed from above to below threshold converts the BEC atoms into
dissociatively stable molecules; i.e., because the dissociation channel opens (closes) in a sweep
beginning below (above) the dissociation threshold. The catch is that the so-formed molecules
are highly vibrationally excited, so that collision-induced relaxation can rather drastically limit
the lifetime of molecular BECs created in this manner.

With this understanding in mind, magnetoassociation experiments with \cs
condensates~\cite{HER03} have recently focused on creating molecules more
efficiently~\cite{MAR04}. For sweeps directed from above to below the threshold for
magnetodissociation, the molecular conversion efficiency was found to saturate at about 10\%
for decreasing sweep rates, which is broadly similar to magnetic-field-sweep experiments with
Rb~\cite{DUE04} and Na~\cite{KXU03}. Faced with this inefficiency,
Mark~\etal~\cite{MAR04} decided to take another tack on making molecules: starting from
well above threshold, they abruptly switch the magnetic field value to its resonance position,
waited a given amount  of time, and then abruptly switched the magnetic field value to a
position well below threshold. Somewhat surprisingly, the result was a three-fold improvement
in the molecular conversion efficiency. Moreover, the improved efficiency was independent of
whether the system was initially above or below threshold, as long as the final switch was to a
magnetic field value well below threshold. Obviously these observations defy an understanding
in terms of adiabatic conversion of atoms into molecules~\cite{DUE04,KXU03}. The purpose
of this Letter is demonstrate that molecular formation in the switch experiments can be
explained as coherent population trapping--Rabi/Josephson oscillations that damp to a
nonzero value--among the coupled atomic condensate, molecular condensate, and
magnetodissociated atoms pairs.

{\it Atom, Molecule, and Pair Model.}--Early theories of collective magnetoassociation
accounted only for the condensates, neglecting any and all noncondensate
modes~\cite{TIM99}. However, rogue~\cite{JAV02,MAC02}, or unwanted~\cite{GOR01},
transitions to noncondensate atom pairs can occur because magnetodissociation of a
zero-momentum BEC molecule need not take the offspring atoms back to the zero-momentum
atomic condensate, but may just as well end up creating two noncondensate atoms with
equal-and-opposite momenta. Since the collective condensate coupling scales like the square
root of the magnetic-field width of the Feshbach resonance and the magnetodissociation rate
scales like the width itself, rogue dissociation is expected to play a dominant role in strong
magnetoassociation. The prime indicator of rogue relevance is the density-dependent frequency
$\omega_\rho=\hbar\rho^{2/3}/m$~\cite{JAV02,MAC02}. When the collective-enhanced
atom-molecule coupling satisfies $\Omega\agt\omega_\rho$, the role of transitions to
noncondensate atom pairs needs to be carefully considered in atom-molecule conversion.

The purpose of this section is to introduce our minimal-yet-realistic
model~\cite{JAV02,MAC02} of a joint atom-molecule condensate undergoing rogue
dissociation to noncondensate atom pairs, in order to determine the mechanism behind the
unexplained molecule conversion in the Innsbruck experiments~\cite{MAR04}. Hence, we
assume that $N$ atoms have Bose-condensed into the same one-particle state, e.g., a plane
wave with wave vector $\bk=0$. Magnetoassociation then removes two atoms from this state
$|1\rangle$, creating a molecule in the molecular state $|2\rangle$, which may then dissociate
into the noncondensate pair state $|\bk,-\bk\rangle$. In second quantized notation, boson
annihilation operators for zero-momentum condensate atoms (molecules) and noncondensate
atoms of momentum $\bk$ are denoted by $a$ ($b$) and $a_\bk$. The field-matter interactions
that drive the atom-molecule transitions are characterized by the Rabi frequency $\kappa$, the
binding energy (detuning) of the molecule is denoted by $\delta$; also, collisions between
condensate particles are included with a coupling strength $\lambda_{ij}$ (more on these
quantities in a moment). The Hamiltonian for this system reads
\begin{widetext}
\beq
\frac{H}{\hbar}=\delta\bdag b -\half\kappa\left[\bdag aa + \adag\adag b\right]
%\nonumber\\&&
  +\half\left[\lambda_{11} \adag\adag aa + \lambda_{22} \bdag\bdag bb\right]
     +\lambda_{12}\adag\bdag ab
%\nonumber\\&&
  +\half\sum_{\bk\neq0} \left[\epsilon_\bk\adag_\bk a_\bk
    -\kappa f_\bk \left(\bdag a_\bk a_{-\bk} +\Hc\right)\right],
\label{HAM}
\eeq
\end{widetext}
where $\epsilon_\bk=\hbar k^2/m$ is the energy of a noncondensate pair, and
$f_\bk$ describes the wavenumber (energy) dependence of the noncondensate
coupling. Collisions with noncondensate atoms are neglected for
simplicity.

The corresponding mean-field theory is derived from the Heisenberg equations of motion:
\bml
\bea
i\dot{a} &=&
  \left[\Lambda_{11}|a|^2+\Lambda_{12}|b|^2\right]a
    - \Omega a^* b, \\
i\dot{b} &=&  \left[\delta+\Lambda_{12} |a|^2 + \Lambda_{22} |b|^2\right] b
\nonumber\\&&
  -\!\half\!\left[\Omega a^2+\xi\!\int\!d\epsilon\sqrt[4]{\epsilon}\,
    f(\epsilon)C(\epsilon)\right]\!\!,\:\:
%\nonumber\\
\\
i\dot{C}(\epsilon) &=&\epsilon C(\epsilon)
    -\xi\sqrt[4]{\epsilon}\,f(\epsilon)\left[1+2P(\epsilon)\right]b,
\\
i\dot{P}(\epsilon) &=&
  \frac{(2\pi\omega_\rho)^{3/2}}{\sqrt[4]{\epsilon}}
    f(\epsilon)\left[b^*C(\epsilon)-C^*(\epsilon)b\right].
\eea
\label{ROGUE_MFE}
\eml
In Eqs.~\eq{ROGUE_MFE}, we have taken the continuum limit
\beq
\frac{1}{N}\sum_\bk G_\bk \rightarrow
  \frac{1}{4\pi^2\omega_\rho^{3/2}}\int d\epsilon\, G(\epsilon),
\eeq
and we also have introduced the rogue and normal densities,
$C(\epsilon)\equiv\sqrt[4]{\epsilon}\,
  \left\langle a_\bk a_{-\bk}\right\rangle/(2\pi\omega_\rho^{3/4})$
and $P(\epsilon)\equiv\langle \adag_\bk a_\bk\rangle$, as well as
the rogue coupling $\xi=\Omega/(2\pi\omega_\rho^{3/4})$. The effects of collective
enhancement have been included by scaling the mean-field amplitudes
according to $x\rightarrow x/\sqrt{N}$ (where $x=a,b$). The mean-field fractions $|a|^2$
and $2|b|^2$ are then of the order of unity, and the couplings have been
redefined as $\Omega=\sqrt{N}\kappa$ and $\Lambda_{ij}=N\lambda_{ij}$.

Finally, the collective-enhanced couplings are given explicitly as
$\Omega=[2\pi\hbar\rho|a_{11}|\Delta_\mu\Delta_B/m^*_{11}]^{1/2}$
and $\Lambda_{ij}=2\pi\hbar\rho a_{ij}/m^*_{ij}$ ($i,j=1,2$), where $a_{11}$ ($a_{12}$,
$a_{22}$) denotes the zero-field atom-atom (atom-molecule, molecule-molecule) scattering
length, $\Delta_\mu$ denotes the difference between the magnetic moments of the free atom
pair and the bound molecule,
$\Delta_B$ denotes the magnetic-field width of the Feshbach resonance, and
$m^*_{ij}$ is the reduced mass of the $ij$-th pair; also, the magnetic field
dependence of the binding energy is given as
$\delta={\rm sgn}[a_{11}]\Delta_\mu(B-B_0)/\hbar$, where $B_0$ is the magnetic-field
position of resonance.

Before moving on, we discuss basic numerics. For $\Omega=\Lambda_{ij}=P=0$ and
$\xi\neq0$, simple Fourier analysis of an initial bound molecular condensate [$b(t=0)=1$] gives
the below-threshold binding energy as the real and negative pole of
$\omega-\delta-\Sigma(\omega)+i\eta=0$, where $\eta=0^+$ and the molecular self-energy is
defined as
\beq
\Sigma(\omega)=\half\xi^2\int
d\epsilon\,f^2(\epsilon)\,
  \frac{\sqrt{\epsilon}}{\omega-\epsilon+i\eta}\,.
\eeq
The simplest energy dependence for the continuum is one that obeys the
Wigner threshold law up to some abrupt cutoff:
$f^2(\epsilon)=\Theta(\epsilon_M-\epsilon)$. The detuning (binding energy) then
picks up a term $\Sigma(0)=\xi^2\sqrt{\epsilon_M}$\,. In principle, the cutoff
is infinite, and therefore so is the continuum shift of the
molecular binding energy. To account for this divergence, one defines the
so-called physical detuning
$\tilde\delta=\delta-\xi^2\sqrt{\epsilon_M}\,$, which is finite by
definition in the limit of an infinite cutoff. In practice, any numerical
procedure employs a finite cutoff, and the finite shift is accounted for
in exactly the same manner, and the intermediate detunings in
Eqs.~\eq{ROGUE_MFE} are taken as physical (renormalized) detunings. The number of
noncondensate quasicontinuum states and cutoff are chosen to deliver convergence and
minimize numerical artifacts.

{\it Results for Sweep and Switch Experiments.}--To enable a discussion of the
\cs sweep and switch experiments, we gather explicit parameters
from Refs.~\cite{HER03,MAR04}: $a_{11}=200a_0$, $\Delta_\mu=0.93\mu_0$,
$\Delta_B=2.1\,$mG, and $B_0=19.84\,$G [where $a_0$ ($\mu_0$) is the Bohr radius
(magneton)]. Although exact values are unknown, we approximate the effect of atom-molecule
and molecule-molecule collisions with $\Lambda_{12}=\Lambda_{22}=\Lambda_{11}$;
similarly, we account for collision-induced vibrational relaxation by borrowing an imaginary
scattering length from Na (see, e.g., Yurovsky~\etal~\cite{TIM99}), so that
$\Im[\Lambda_{12}]=\Im[\Lambda_{22}]=\rho\times 10^{-11}\,{\rm s}^{-1}$. Crucially,
$\Omega/\omega_\rho\sim 2$ for $\rho\sim 10^{12}\,{\rm cm}^{-3}$, so that rogue
dissociation to noncondensate modes must be treated carefully, despite the narrow
resonance~\cite{ROGIES}.

In order to ensure a uniform magnetic field across the atomic sample, the
combined optical-dipole and magnetic-field-levitation trap is turned off, leading to ballistic
expansion over the course of the experiments~\cite{MAR04}. Nevertheless, we simply assume
a peak density consistent with $\sim 10\,$ms of ballistic expansion,
$\rho_0=1.1\times10^{12}\,{\rm cm}^{-3}$~\cite{MAR04}, and account for the initial
inhomogeneity with a local density approximation. This overestimates the
density and, in turn, the role of vibrational quenching; on the other hand, the
magnetic-field range here is truncated, and the sweep rate decreased, which ultimately shortens
the total sweep time and so underestimates relaxational losses. Of course, these two factors
do not necessarily cancel each other out, and the ensuing results are only qualitative; the aim is
not to quantitatively model the sweep experiments anyway, but to explain the improved
efficiency of the switch experiments.

\begin{figure}
\centering
\includegraphics[width=8.0cm]{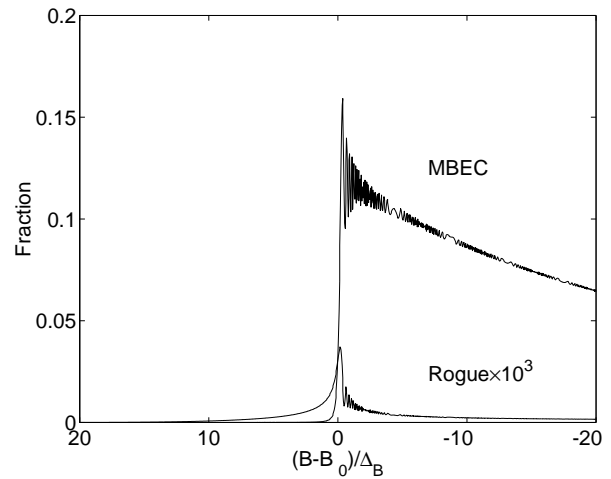}
\caption{Coherent conversion of a \cs atomic condensate into a \csm condensate by
sweeping a magnetic field across a Feshbach resonance. Here $\dot{B}=1\,$G/s, making for a
total sweep time of $84\,$ms, which is, evidently, plenty of time for vibrational relaxation to
filch $\sim5\%$ of the MBEC population.}
\label{SWEEP}
\end{figure}

In the sweep experiments~\cite{MAR04}, the magnetic field is initially tuned
0.5~G above the Feshbach resonance threshold, $B_i=B_0+0.5\,$G, which is ramped in a linear
fashion, $B(t)=B_0-\dot{B}t$, to a final value 0.5~G below threshold, $B_f=B_0-0.5\,$G. For
experimental ramps with $\dot{B}\alt 10\,$G/s the total sweep time is $\agt100\,$ms, and the
conversion efficiency saturates at ca. 10\%. For numerical ease the B-field range here is
truncated to $B=B_0\pm 20\Delta_B$, and we use $\dot{B}=1\,$G/s, for a total sweep
time of $84\,$ms. The results are shown in Fig.~\ref{SWEEP}. The agreement is
surprisingly good, all things considered. Also, as the noncondensate population begins to build,
it is transferred into the molecular BEC; hence, creation of molecules instead of dissociated
pairs is more about coherent conversion than a closing magnetodissociation channel (the effect
is even more dramatic for slower sweeps, as per Ref.~\cite{MAC02b}).

The relevant frequency scale is $\Omega$~\cite{MAC02,JAV02}, so that adiabatic following
requires $\dot\delta/\Omega\alt 1$. For $\rho_0$ and $\dot{B}$ as above,
$\dot\delta/\Omega_0\approx 8$; hence, the center of the cloud is still on the outside of the
adiabatic regime, and the situation is worse on the edges of the cloud. Slower sweeps therefore
lead to more molecules, but allow more time for vibrational relaxation, leading to a saturation of
the molecular conversion efficiency. Nevertheless, it is only necessary to satisfy adiabaticity in
the immediate vicinity resonance, so it should be worth trying to engineer a ramp that is
``ultrafast" except in the region $B\sim B_0\pm\Delta_B$.
In this manner, adiabaticity could be satisfied in the vicinity of resonance, while minimizing the
total sweep time and, thus, vibrational relaxation losses, yielding more MBEC.

\begin{figure}
\centering
\includegraphics[width=8.0cm]{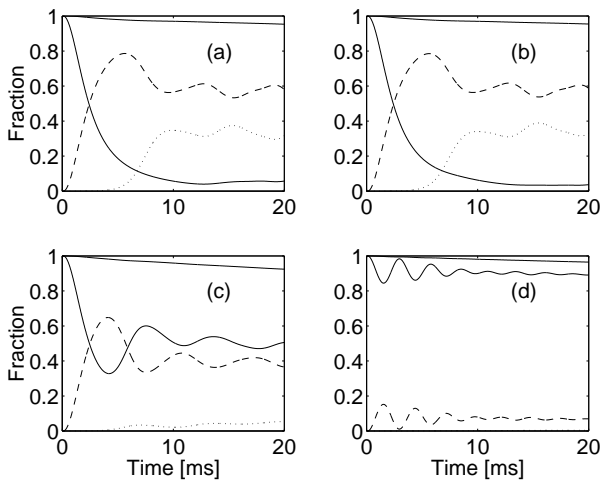}
\caption{Coherent population trapping in a Feshbach-resonant \cs Bose-Einstein condensate,
following a sudden ($0.1\,$ms) switch to the vicinity of threshold. In each panel, the total
fraction is the uppermost (linearly decreasing) solid line, whereas the BEC (MBEC,
rogue/noncondensate) fraction is given by the oscillating solid (dashed, dotted) line. Population
trapping is independent of whether the system is initially above (a) or below (b) threshold.
However, the amount of population trapped does depend on the final magnetic field, as
illustrated by (c) $B_f=B_0+\Delta_B/100$ and (d) $B_f=B_0+\Delta_B/10$, although it does
not really matter whether $B_f$ is above or below threshold (not shown). The rogue fraction in
panel (d) is ca. $10^{-3}$ and is not visible.}
\label{SWITCH}
\end{figure}

Now we turn to the Ref.~\cite{MAR04} switch experiments. Here
the magnetic field is suddenly switched to $B\sim B_0$ in a time $\sim10\,$ms, held for a
given amount of time, and then suddenly switched to a well-below-threshold value where the
fraction of molecules is determined. Ballistic expansion is not really an issue when modeling
the switch experiments: the switch occurs in a time $\sim10\,$ms, and the hold times are
$\sim10\,$ms themselves, so it is entirely reasonable to use $\rho_0=1.1\times10^{12}\,{\rm
cm}^{-3}$ throughout; moreover, the switch time here is shortened, without changing the
physics, to hasten the numerics. The results are shown in Fig.~\ref{SWITCH}. Whereas not
much happens during the switch period, during the hold period the system undergoes
coherent Josephson/Rabi oscillations between the atomic condensate, the molecular
condensate, and noncondensate atom pairs; however, while the couplings remain, the
system subsequently freezes into a superposition immune to magnetodissociation. The sudden
switch back then harvests the frozen population. The independence of the whether the initial
$B$-field is above or below threshold is seen by comparing Figs.~(a,b). More importantly, an
intriguing feature of the experiments is an apparent step-wise threshold in the conversion
efficiency at $\sim15\,$ms, beyond which the molecular efficiency is $\sim30\%$. Due to an
inadvertent lag~\cite{NAG04}, the $B$-field value does not reach the vicinity of resonance as
expected, so the system is still off resonance; along with the fact that trapped-population
fraction is depends on the detuning, as per Figs.~(c,d), the lag leads to an apparent threshold.
Similarly, experimental fluctuations in the magnetic field value~\cite{MAR04} could easily
account for the difference between the observed ($\sim30\%$) and the calculated
($\sim60\%$) molecular fraction for $B_f=B_0$.

{\it Conclusions.}--Given that faster sweeps apparently convert fewer an fewer atoms into
molecules, improving the molecular conversion efficiency by {\em speeding up} the sweep is
truly a surprise. The plot thickens with the observation of a final \csm fraction that is
independent of whether the initial magnetic field is tuned above or below threshold (as long as
the final switch is to below threshold), in contrast with previous
experiments~\cite{DUE04,KXU03}. Nevertheless, it is known from past work on
photodissociation of a negative ion~\cite{RZA82} that--given a shaped continuum--some of the
population may stay permanently trapped in the bound state. Indeed, the same effect was
predicted for strong photoassociation~\cite{JAV02}, and since the theory of a photoassociation
and magnetoassociation resonance are formally equivalent, the answer is obvious in hindsight.
And yet it is still surprising that, given all the possible decoherence avenues, coherent
population trapping could survive. The results herein strongly suggest that it does, making
coherent population trapping the leading candidate to explain how a fast switch
could get the better of a slow sweep.

{\it Acknowledgements.}--Cheng Chin, Rudi Grimm, Christoph N\"agerl for helpful
conversations; express thanks to Marwan Rasamny for helpful conversations and use of
Delaware State University computational resources.


\begin{thebibliography}{99}
\bibitem{ROGIES}
  Because $\Omega\propto(a_{11}\Delta_B)^{1/2}$,
not just $\Delta_B^{1/2}$; the low density also plays a role, albeit weak
($\Omega/\omega_\rho\propto\rho^{-1/6}$).
\end{thebibliography}
\end{document}